\documentclass[12pt, a4paper, twocolumn]{article}
\usepackage{xurl}
\usepackage{abstract}

\usepackage{geometry}
\usepackage{times}
\usepackage{epsfig}
\usepackage{graphicx}
\usepackage{amsmath}
\usepackage{amssymb}

\usepackage{caption}
\usepackage{subcaption}
\usepackage{booktabs}
\usepackage{authblk}
\usepackage{stfloats}

\usepackage{hyperref}
\hypersetup{colorlinks=true, urlcolor=blue, linkcolor=blue, citecolor=blue}

\title{Deep Diffusion Models for Seismic Processing}

\author[1,2]{Ricard Durall}
\author[1,2]{Ammar Ghanim}
\author[1,2,3]{Mario Fernandez}
\author[1,2]{Norman Ettrich}
\author[1,2,4]{Janis Keuper}
\affil[1]{Fraunhofer ITWM}
\affil[2]{Fraunhofer Center Machine Learning}
\affil[3]{École Normale Supérieure}
\affil[4]{IMLA, Offenburg University}

\date{}

\newcommand{\abstractText}{\noindent
Seismic data processing involves techniques to deal with undesired effects that occur during acquisition and pre-processing.
These effects mainly comprise coherent artefacts such as multiples, non-coherent signals such as electrical noise, and loss of signal information at the receivers that leads to incomplete traces.
In the past years, there has been a remarkable increase of machine-learning-based solutions that have addressed the aforementioned issues.
In particular, deep-learning practitioners have usually relied on heavily fine-tuned, customized discriminative algorithms.
Although, these methods can provide solid results, they seem to lack semantic understanding of the provided data.
Motivated by this limitation, in this work, we employ a generative solution, as it can explicitly model complex data distributions and hence, yield to a better decision-making process.
In particular, we introduce diffusion models for three seismic applications: demultiple, denoising and interpolation.
To that end, we run experiments on synthetic and on real data, and we compare the diffusion performance with standardized algorithms.
We believe that our pioneer study not only demonstrates the capability of diffusion models, but also opens the door to future research to integrate generative models in seismic workflows.
}

\begin{document}

%%%%%%%%%%%%
% Abstract %
%%%%%%%%%%%%

\twocolumn[
  \begin{@twocolumnfalse}
    \maketitle
    \begin{abstract}
      \abstractText
      \newline
      \newline
    \end{abstract}
  \end{@twocolumnfalse}
]

%%%%%%%%%%%
% Article %
%%%%%%%%%%%

\section{Introduction}

Deep generative learning has become an important research area in the machine learning community, being more relevant in many applications.
Namely, they are widely used for image synthesis and various image-processing tasks such as editing, interpolation, colourization, denoising, and super-resolution.
Recently, diffusion probabilistic models \cite{sohl2015deep,ho2020denoising} have emerged as a novel, powerful class of generative learning methods.
In a short period of time, these models have achieved surprisingly high performance \cite{dhariwal2021diffusion,saharia2021palette,rombach2022high, ramesh2022hierarchical}, and have even surpassed state-of-the-art algorithms like generative adversarial networks \cite{goodfellow2014generative} (GANs) and variational autoencoders \cite{kingma2013auto} (VAEs).

At the same time, the geophysics community has been actively adopting deep-learning techniques to boost and automate numerous seismic interpretation tasks including fault picking \cite{an2021deep,wu2019faultSeg}, salt delineation \cite{oh2018salt,shi2019saltseg}, well-to-seismic tie \cite{nivlet2020automated,tschannen2022partial}, horizon tracking \cite{yang2020seismic,tschannen2020extracting}, multiple removal \cite{bugge2021demonstrating,durall2022dissecting}, etc.
Nonetheless, to the best of our knowledge, there has not been yet any work exploring the application of diffusion models to seismic data and thus, studying their potential advantages to already established deep-learning approaches in this domain.
Driven by this motivation, in this work, we study the applicability of diffusion models for seismic processing.

Seismic imaging is essential to discover and characterize economically worthwhile geological reservoirs, such as hydrocarbons accumulations, and to manage the extraction of the resources stored in them.
Unfortunately, recorded seismic signals at the surface are inevitably contaminated by coherent and incoherent noise of various nature.
The process of removing the noise, while retaining the primary signal, is called seismic processing.
In this paper, we focus on three relevant, well-known seismic processing tasks: demultiple, denoising and interpolation.
Demultiple and denoising are both removing unwanted signals from the seismic section; the first gets rid of coherent noise caused by reverberations of waves between strong reflectors, whereas the latter removes incoherent noise of miscellaneous causes.
The goal of interpolation is to fill-in gaps in the image caused by limitations during acquisition. 
Although at the first glance the nature of these problems might look different or unrelated, it is possible to formulate a common framework, in which they can be solved.
This is feasible, due to the fact that the diffusion models, like most of generative models, learn the density distribution of the input data.
In other words, unlike discriminative approaches which draw boundaries in the data space, the generative approaches model how data is placed throughout the space \cite{tomczak2022deep}.
As a result, they are powerful algorithms that can be independently applied to a large diversity of problems.

\section{Background}
\label{appendix}
Generative models for modelling estimate the marginal distribution, denoted as $p(x)$, over observable variables $x$, e.g., images.
In the literature, we can find different formulations that tackle this problem such as autoregressive generative models, latent variable models, flow-based models, and energy-based models.

\subsection{Latent Variable Models}
The main idea of this type of models is to utilize latent variables $z$ to formulate the joint distribution $p(x,z)$, which describes the marginal distribution as a function of learnable parameters ${\theta}$ (likelihood). 
Mathematically, it can be written as:
\begin{equation}
    \begin{split}
	& z \sim p_{\theta}(z) \\
	& x \sim p_{\theta}(x|z) \\
	p_{\theta}(x) = \int_z p_{\theta}(x &,z) = \int_z p_{\theta}(x|z) p_{\theta}(z).
	\end{split}
\end{equation}

Unfortunately, for most of the problems we do not have access to the true distribution $p(x)$ and hence, we need to fit our model to some empirically observed subset.
One solution is to use Monte Carlo sampling to approximate the integral over $z$ to try to estimate the model parameters $\theta$.
Nonetheless, this approach does not scale to high dimensions of $z$ and consequently, we will suffer from issues associated with the curse of dimensionality.
Another solution is to use variational inference, e.g., VAE \cite{kingma2013auto}.
In particular, the lower bound of the log-likelihood function, called the Evidence Lower BOund (ELBO).
The ELBO provides a joint optimization objective, which simultaneously updates the variational posterior $q_{\phi}(z|x)$ and likelihood model $p_{\theta}(x|z)$.
The objective is written as:
\begin{equation}
    \begin{split}
    \mathrm{log} \, p(x) & \geq \mathbb{E}_{z\sim q_{\phi}(z|x)}[\mathrm{log} \, p_{\theta}(x|z)] \\
    & -\mathrm{KL}[q_{\phi}(z|x)||p(z)],
    \end{split}
\end{equation}
where KL stands for the Kullback-Leibler divergence.

\begin{figure}
    \begin{subfigure}{.5\textwidth}
        \centering
        \includegraphics[width=.4\linewidth]{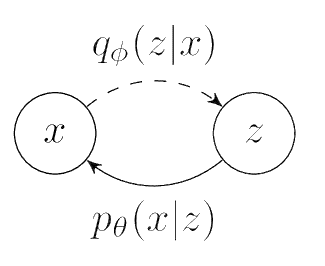}
    \end{subfigure}\\[1.5ex]
    \begin{subfigure}{.5\textwidth}
        \centering
        \includegraphics[width=.65\linewidth]{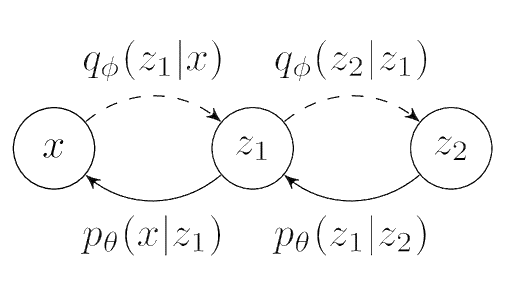}
    \end{subfigure}\\[1.5ex]
    \begin{subfigure}{.5\textwidth}
        \centering
        \includegraphics[width=.9\linewidth]{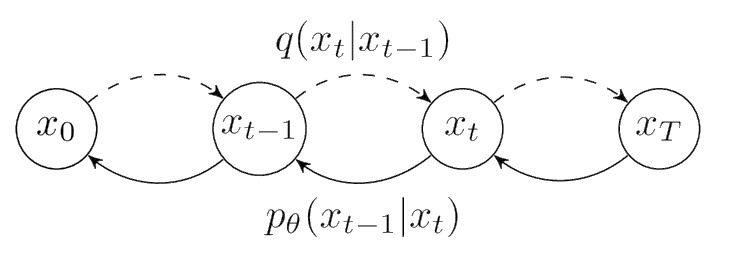}
    \end{subfigure}
    \caption{ Scheme of the different latent variable models.
    (Top) Single latent variable model.
    (Center) Hierarchical latent variable model.
    (Bottom) Diffusion model.}
    \label{fig:scheme}
\end{figure}

\subsection{Hierarchical Latent Variable Models}
Once defined a single stochastic layer, it is straightforward to derive hierarchical extensions. 
For example, let us consider a latent variable model with two latent variables $z_1$ and $z_2$.
We can define the joint distribution $p(x,z_1,z_2$) and marginalizing out the latent variables:
\begin{equation}
    \begin{split}
    p_{\theta}(x) = &\int_{z_1} \int_{z_2} p_{\theta}(x,z_1,z_2) \\
    = &\int_{z_1} \int_{z_2} p_{\theta}(x|z_1)p_{\theta}(z_1|z_2)p_{\theta}(z_2).
    \end{split}
\end{equation}
Similar to the single latent model, we can derive the variational approximation (ELBO) to the true posterior as:
\begin{equation}
    \begin{split}
    \mathrm{log} \, p(x) & \geq \mathbb{E}_{z_1\sim q_{\phi}(z_1|z_2)}[\mathrm{log} \, p_{\theta}(x|z_1)] \\
    & -\mathrm{KL}[q_{\phi}(z_1|x)||p_{\theta}(z_1|x)] \\
    & -\mathrm{KL}[q_{\phi}(z_2|z_1)||p(z_2)]. 
    \end{split}
\end{equation}

\subsection{Diffusion Models}
Diffusion models belong to the latent variable family as well.
In fact, we can think of them as a specific realization of a hierarchical latent variable model, where the inference model\footnote{Remember that the inference model relates a set of observable variables to a set of latent variables, e.g., $q(z|x)$.} does not have learnable parameters.
Instead, it is constructed so that the final latent distribution $q(x_T)$ converges to a standard Gaussian (where $T$ is the number of latent variables).
The objective function of diffusion models is written as:
\begin{equation}
\begin{gathered}
    \mathrm{log} \, p(x) \geq  \\
     \mathbb{E}_{x_{1:T} \sim q(x_{1:T}|x_0)} [\mathrm{KL} (q(x_{T}|x_0)||p_{\theta}(x_{T})) \\
    + \sum_{t=2}^{T} \mathrm{KL} (q(x_{t-1}|x_t,x_0)||p_{\theta}(x_{t-1}|x_t))\\
    - \mathrm{log} \, p_{\theta}(x_{0}|x_1)].
\end{gathered}
\end{equation}
Under certain assumptions, this objective can be further simplified, leading to the following approximation:
\begin{equation}
    \begin{split}
    \mathrm{log} \, p(x)  &\gtrsim \sum_{t=2}^{T} \mathrm{KL} (q(x_{t-1}|x_t,x_0)||p_{\theta}(x_{t-1}|x_t)) \\
    & = \sum_{t=2}^{T} || \epsilon - \epsilon_{\theta}(\sqrt{\Bar{\alpha}_t}x_0+\sqrt{1-\Bar{\alpha}_t}\epsilon,t)||^2.
    %\sum_{t=2}^{T} ||\Tilde{\mu}_t(x_t,x_0) - \mu_{\theta}(x_t,t)||^2
    \end{split}
\end{equation}
%where $\Tilde{\mu}_t(x_t,x_0)$ is a linear combination of $x_t$ and $x_0$ that depends on the variance schedule $\beta_t$.
Note that we drop the expectation for clarity.
The exact derivation can be found in \cite{ho2020denoising}.

\section{Methodology}
In this section, we provide a brief overview of diffusion models formulation.
Note that we do not aim at covering the entire derivations.
For a more in-depth, detailed mathematical description, we refer the reader to \cite{ho2020denoising}. 

\subsection{Background}
On a high level, diffusion models consist of two parts: forward diffusion and parametrized reverse.
The forward diffusion part can be described as a process, where Gaussian noise $\epsilon$ is gradually applied to the input image $x_0$ until the image becomes entirely unrecognizable from a normal distribution $x_T \sim \mathcal{N}(0,\mathrm{I})$ ($T$ is the number of transformation steps).
That is to say, at each step of this process, the noise is incrementally added to the data,  $x_0 \xrightarrow{+\epsilon} x_1 \xrightarrow{+\epsilon} ... \xrightarrow{+\epsilon} x_T$.
This procedure together with the Markov assumption\footnote{Markov assumption is used to describe a model that holds the memoryless property of a stochastic process.} leads to a simple parameterization forward process expressed as:
\begin{equation}
    \begin{split}
    q(x_{1:T}|x_0) & = \prod_{t = 1}^{T} q(x_t|x_{t-1}) \\
    & = \prod_{t = 1}^{T} \mathcal{N}(x_t; \sqrt{1-\beta_t}x_{t-1},\beta_t \mathrm{I}),
    \end{split}
    \label{eqn:forward}
\end{equation}
where the variable $\beta$ defines a fixed variance schedule, chosen such that $q(x_T|x_0) \approx \mathcal{N}(0,\mathrm{I})$.

\begin{figure}[t]
  \centering
  \includegraphics[width=.9\linewidth]{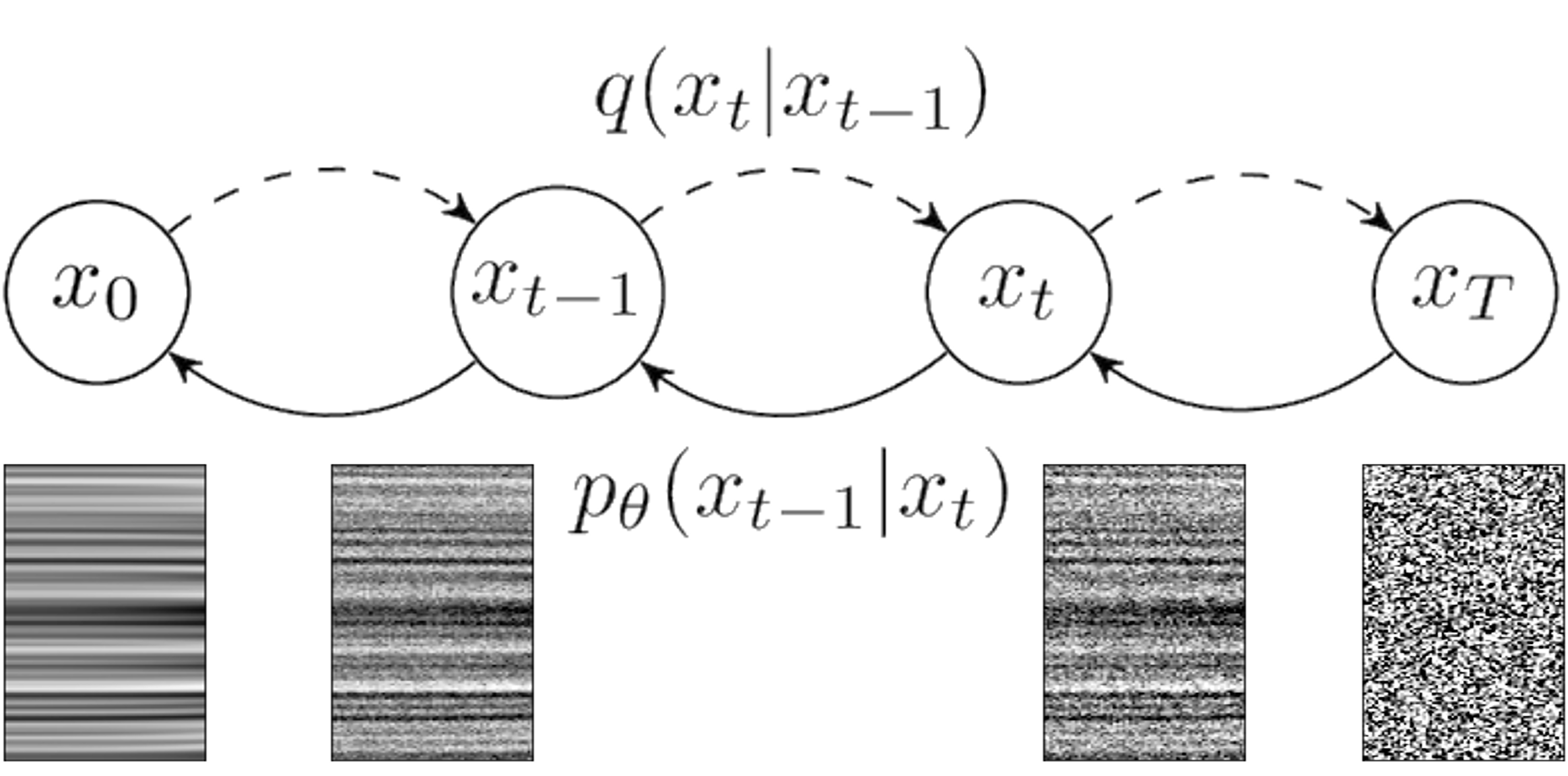}
  \caption{Denoising diffusion process.
  While the Markov chain of the forward diffusion gradually adds noise to the input (dash arrows), the reverse process removes it stepwise (solid arrows). }
  \label{fig:main}
\end{figure}

The second part, the parametrized reverse process, represents the data synthesis.
Thus, it undoes the forward diffusion process and performs iterative denoising.
To that end, the reverse process is trained to generate data by converting random noise into realistic data.
Formally, this generative process is defined as a stochastic process, which iteratively removes noise from the input images using deep neural networks.
Starting with the pure Gaussian noise $p(x_T) = \mathcal{N}(x_T, 0,\mathrm{I})$, the model learns the joint distribution $p_{\theta}(x_{0:T})$ as:
\begin{equation}
    \begin{split}
    p_{\theta}(x_{0:T}) & = p(x_T) \prod_{t = 1}^{T} p_{\theta}(x_{t-1}|x_t) \\
    & = p(x_T) \prod_{t = 1}^{T} \mathcal{N}(x_{t-1};\mu_{\theta}(x_t,t), \Sigma_{\theta}(x_t,t)),
    \end{split}
    \label{eqn:reverse}
\end{equation}
where the time-dependent parameters of the Gaussian transformations $\theta$ are learned.
Note in particular that the Markov formulation asserts that a given reverse diffusion transformation distribution depends only on the previous timestep.

\subsection{Training}
A diffusion model is trained by finding the reverse Markov transitions that maximize the likelihood of the training data.
In practice, this process consists of optimizing the variational lower bound on the log likelihood.
Hereunder the simplified expression derived by \cite{ho2020denoising}:
\begin{equation}
    \begin{split}
    \mathrm{log} \, p(x)  &\gtrsim  \sum_{t=2}^{T} || \epsilon - \epsilon_{\theta}(\sqrt{\Bar{\alpha}_t}x_0+\sqrt{1-\Bar{\alpha}_t}\epsilon,t)||^2,
    \end{split}
\end{equation}
where
\begin{gather}
    \alpha_t = 1- \beta_t \; \mathrm{and} \; \Bar{\alpha}_t = \prod_{i = 1}^{T} \alpha_i.
\end{gather}
Note, ultimately, the deep neural network learns to predict the noise component $\epsilon$ at any given timestep.

\section{Experiments}
In this section, we validate the flexibility of diffusion models for different seismic tasks.
In particular, we analyse three case studies: demultiple, denoising and interpolation.
To do that, we present an end-to-end deep-learning approach that can deal (separately) with demultiple, denoising and interpolation scenarios.
Furthermore, we benchmark the results with alternative paradigms that are currently employed in both academia and industry domains.

The implementation details are as following: In all our experiment, we train the diffusion model for 200,000 iterations with a batch size of 32; we set $\beta$ to follow a linear schedule, and we use a depth of 2000 timesteps for both the forward process (see Equation \ref{eqn:forward}) and the reverse denoising process (see Equation \ref{eqn:reverse}).

\begin{figure}[t]
  \centering
  \includegraphics[width=\linewidth]{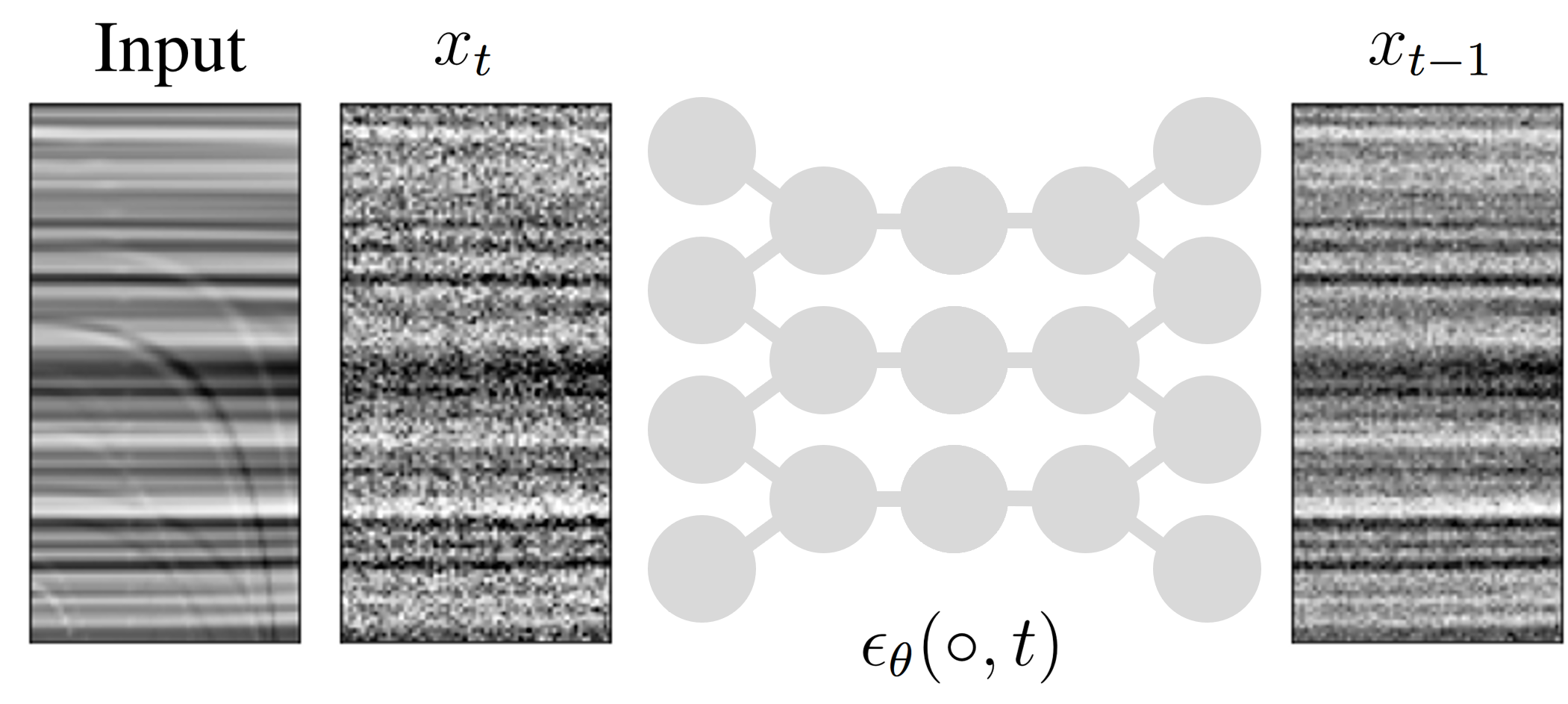}
  \caption{In each reverse step $t$, the model  $\epsilon_{\theta}$ is fed with the semi-denoised multiple-free image $x_t$ and the multiple-infested input.
  As an output, the network generates the image $x_{t-1}$, which should have less noise and no multiples. }
  \label{fig:demultiple_model}
\end{figure}

\begin{figure*}[b]
    \begin{subfigure}{\textwidth}
        \centering
        \includegraphics[width=\linewidth]{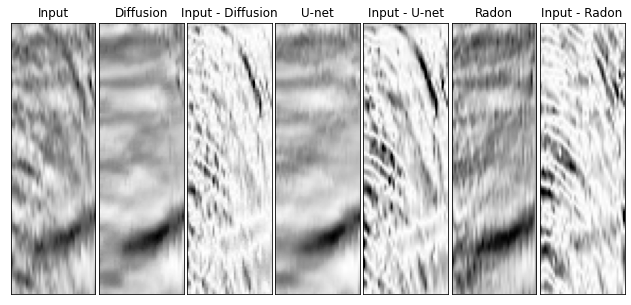}
    \end{subfigure}\\[-1ex]
    \caption{This figure displays two cropped gathers that contain multiples (input), and the results after applying the demultiple algorithms.
    Moreover, we plot the difference between the input and the output to check the content that has been removed.
    Note that we apply a scaling factor of 3 in the differences to stress the changes.}
    \label{fig:demultiple_results}
\end{figure*}

\subsection{Architecture}
Image diffusion models commonly employ a time-conditional U-net \cite{ronneberger2015u}, parametrized as $\epsilon_{\theta}(\circ, t)$, as a neural backbone.
This architecture was initially introduced in \cite{ho2020denoising}, where the main motivation for this topology choice was the requirement for the model to have identical input and output dimensionality.
The architecture consists of a stack of residual layers and downsampling convolutions, followed by a stack of residual layers with upsampling convolutions; skip connections connect the layers with the same spatial size.
Furthermore, it uses a global attention layer with a single head to add a projection of the timestep embedding into each residual block.

\subsection{Demultiple}
Primary seismic reflections are events which have reflected only once, and they are employed to describe the subsurface interfaces.
Multiples, on the contrary, are events which appear when the signal has not taken a direct path from the source to the receiver after reflecting on a subsurface boundary.
The presence of multiples in a recorded dataset can trigger erroneous interpretations, since they do not only interfere with the analysis in the post-stack domain, e.g., stratigraphic interpretation, but also with the analysis in the pre-stack domain, e.g., amplitude variation with offset inversion.
Thereby, the demultiple process plays a crucial role in any seismic processing workflow.

In this first experiment, we follow the approach from \cite{breuer2020deep,durall2022dissecting}, and generate synthetic pairs of multiple-infested and multiple-free gathers.
This data setup allows us to train the model in a supervised manner and therefore, we can frame the demultiple problem as an image-to-image transformation task, where the network learns to remove the multiples without removing primary energy.
As in \cite{durall2022dissecting}, the training dataset is designed to include a rich amount of features present in real datasets, to maximize transferability to real case uses.
To that end, we employ as a baseline a conditional diffusion models proposed by \cite{saharia2021image}.
More specifically, we condition our model by concatenating the semi-denoised multiple-free image $x_t$ with the multiple-infested input (see Figure \ref{fig:demultiple_model}).
Ideally, the network should return an improved semi-denoised multiple-free gather $x_{t-1}$ that after $T$ reverse steps should converge into a noise- and multiple-free gather $x_0$.

Once the model is trained, it is crucial to assess the inference capabilities of the network when working on real data, i.e., generalizability.
Nonetheless, this is not a granted property in deep-learning models due to the distribution gap between different datasets, e.g., the gap between synthetic and real datasets \cite{durall2020synthesizing}.
In our experiments, we test the diffusion approach on the dataset from the Volve field made available under Equinor Open Data Licence.
Furthermore, we compare the outcomes with two other multiple-attenuation methodologies: one based on Radon-transform \cite{hampson1986inverse} and one based on deep learning \cite{durall2022dissecting}.
Figure \ref{fig:demultiple_results} shows an example of such a comparison, where we can observe how the diffusion solution offers competitive results, despite minimal hyperparameter tuning involved.
For additional results, see Figure \ref{fig:demo_evo} in the Appendix.

\begin{figure}[b!]
  \centering
  \includegraphics[width=\linewidth]{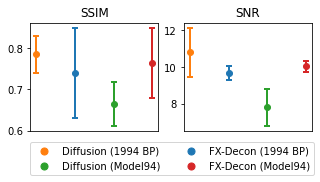}
  \caption{Mean and standard deviation of SSIM and SNR metrics calculated on 500 random denoised images.
  Results from diffusion and FX-Decon scenarios. }
  \label{fig:denoising_metrics}
\end{figure}

\begin{figure}[t!]
  \centering
  \includegraphics[width=\linewidth]{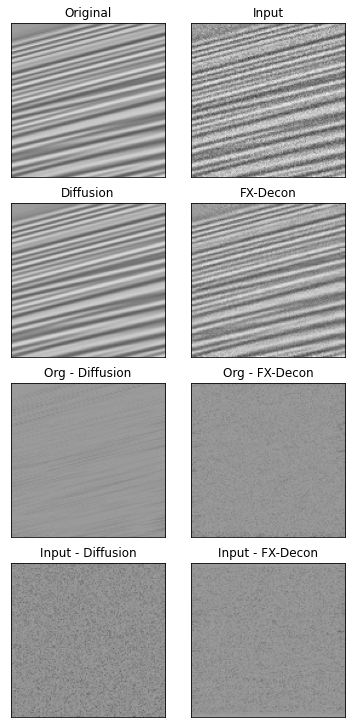}
  \caption{This figure shows an example of denoising.
        The first row contains the original image and the input image (original with noise).
        The second row presents the diffusion and the FX-Decon results.
        Finally, the third and four rows display the difference between the results and the original and the input data, respectively.}
  \label{fig:denoising_results}
\end{figure}

\begin{figure*}[t]
  \centering
  \includegraphics[width=\linewidth]{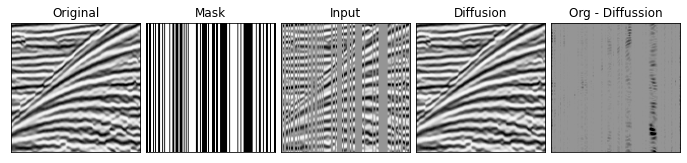}
  \caption{This figure shows an example of interpolation.
        From left to right: the original image, the mask, the input image (original with mask), the diffusion result and its difference with respect to the original image.}
  \label{fig:interpolation_example}
\end{figure*}

\subsection{Denoising}
Incoherent noise can be caused by superposition of numerous unwanted signals from various sources such as ocean waves, wind and electrical instrument noise among others.
Removing such incoherent noise can improve the overall signal-to-noise ratio and, consequently, increase the certainty of interpretation. 
Traditional approaches can be subdivided into two main categories: the prediction filtering methods and domain transform methods.
The first type assumes linearity and predictability of the signal, and constructs a predictive filter to suppress the noise \cite{gulunay1986fx,galbraith1991random}. 
These methods have been widely adopted by the industry due to their efficiency, although they tend to under-suppress noise and occasionally suffer from signal leakage \cite{gulunay2017signal}.
The second type of methods uses mathematical transformations, e.g., Fourier transform \cite{naghizadeh2012seismic}, wavelet transform \cite{mousavi2016automatic}, curvelet transform \cite{neelamani2008coherent,herrmann2008curvelet}, to steer the seismic data into domains, where seismic signals and noise can be easier separated and then leverage the sparse characteristics of seismic data.
This approach, however, often requires a time-consuming transform coefficient tuning.
To cope with this drawback, a new trend based on deep-learning algorithms has emerged, resulting in optimized solutions that remove incoherent noise from seismic data as well as speed up the inference time \cite{yu2019deep,saad2020deep}.

Similar to the demultiple scenario, we create pairs of images to train our diffusion model.
Nonetheless, this time, the objective is to eliminate undesired uncorrelated noise, while preserving the inherent characteristics of the data.
To that end, the pairs of training data consist of a real image and their noisy version.
To create the noisy images, we synthetically add Gaussian noise to the original real images with a variability of the 50\% of their energy.
For this second case of study, we train on 1994 BP \cite{gray1995migration} dataset, from which we extract random patches (from different shot gathers) that neither overlap among each other, nor have more than 40\% of their content equal to 0.
In this fashion, we try to guarantee certain level of variety in the training data.

For the testing set, we apply the same conditions as for training.
Additionally, we employ a second dataset (Model94 \cite{gray1995migration}) to evaluate the generalization capacity of our system.
As for comparison, we use a spectral filtering technique based on the Fourier transform, namely a complex Wiener prediction filter called FX-Decon \cite{gulunay1986fx,galbraith1991random}, which is dedicated for signal extraction and non-coherent noise suppression in the frequency domain.
To assess the results, we use structural similarity index (SSIM) and signal-to-noise ratio (SNR) as quantitative metrics.
Figure \ref{fig:denoising_metrics} displays them for each configuration, i.e., different datasets and methods, and we can observe how the diffusion model provides the best scores when we test on data coming from the same dataset as the one used for training.
However, as expected, it has a drastic drop when we test on a new dataset, e.g., Model94.
This phenomenon is mainly caused by the distribution gap between different datasets.
On the other hand, FX-Decon achieves similar performance on both datasets (no drop), as this method does not involve any learning, i.e., data fitting.
Finally, Figure \ref{fig:denoising_results} illustrates a denoising example for both algorithms.
The difference between the outputs and the original data (third row in Figure \ref{fig:denoising_results}) allows us to see that diffusion model removes some coherent signal, while FX-Decon does not.
Ideally, this should be corrected, but we leave this improvement for future work.
Nevertheless, overall, the diffusion approach leads to less noisy outputs, as can be noticed in the output image.
For additional results, see Figure \ref{fig:denoi_evo} in the Appendix.

\subsection{Interpolation}
Seismic data processing algorithms greatly benefit
from regularly sampled and reliable data.
However, it is rarely the case where the acquired data is presented flawless, i.e., complete shot gathers without missing traces.
Frequently, the reason for that are acquisition constraints such as geophones issues, topography, and economical limitations.
As a consequence, interpolation techniques are a fundamental key for most seismic processing systems.

\begin{figure}[t!]
  \centering
  \includegraphics[width=\linewidth]{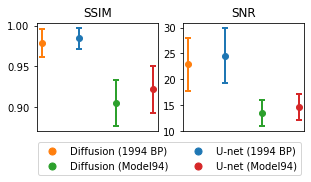}
  \caption{Mean and standard deviation of SSIM and SNR metrics calculated on 500 random interpolated images.
  Results from diffusion and U-net scenarios.}
  \label{fig:interpolation_metrics}
\end{figure}

In this last case of study, we evaluate the capacity of our diffusion model to interpolate missing traces.
To that end, we follow the evaluation methodology introduced by \cite{fernandez2022comparison}, namely, we consider the scenario with irregular missing traces and with a level of decimation set to 50\% (see Figure \ref{fig:interpolation_example}).
Regarding the data for this experiment, we repeat the setup presented in the denoising section, using 1994 BP dataset for training and testing, and Model94 for testing on a new dataset.
Finally, to have a baseline to compare with, we implement the so-called ``standard'' topology from \cite{fernandez2022comparison}, which is essentially a U-net-like network.

Figure \ref{fig:interpolation_metrics} shows the qualitative evaluation of the diffusion approach and of the U-net baseline.
Although results from the latter are superior, the improvement could be considered marginal given the small metric differences.
Furthermore, both algorithms seem to struggle when inferring on unseen datasets.
On the other hand, besides the quantitative results, the potential that diffusion models might bring is objectively higher than discriminative models, as the former are generative models and therefore, can capture more advanced data properties. 
For additional results, see Figure \ref{fig:inter_evo} in the Appendix.

\section{Discussion}
In this work, we propose a generative framework based on diffusion models to address several seismic tasks.
In particular, our case studies include demultiple, denoising and interpolation.
To solve them, we define the problem as an image-to-image transformation, where we have an input image that requires certain modifications so that, the output result belongs to the target domain.
For example, in the demultiple scenario, given a multiple-infested
gather (input domain), our diffusion approach has to identify the multiples and cancel them out, leading to a multiple-free output gather (target domain).

The results of our experimental evaluations are fairly encouraging, as they show competitive performance, when comparing with standardized, customized algorithms.
As we pointed out before, diffusion models for seismic data is an unexplored field to date and hence, the ultimate goal of this project is not to outperform these current algorithms in their respective areas, but to provide a solid analysis of the applicability and flexibility of this novel framework.
Therefore, the main success of our implementation can be regarded as proof of concept that can be used to adopt generative models, namely diffusion models, in the geoscience community.
We believe that our work can help to lay the foundation for future research that can benefit both academia and industry.

\section{Acknowledgement}
This work was developed in the Fraunhofer Cluster of Excellence Cognitive Internet Technologies.
The authors would like to acknowledge the members of the Fraunhofer ITWM DLSeis consortium (http://dlseis.org) for their financial support.
We are also grateful to Equinor and Volve Licence partners for releasing Volve seismic field data under an Equinor Open Data Licence.

\bibliographystyle{ieeetr}
\bibliography{biblio}

\section{Appendix}
\subsection{More Results}
\begin{figure*}[h]
    \begin{subfigure}{\textwidth}
        \centering
        \includegraphics[width=\linewidth]{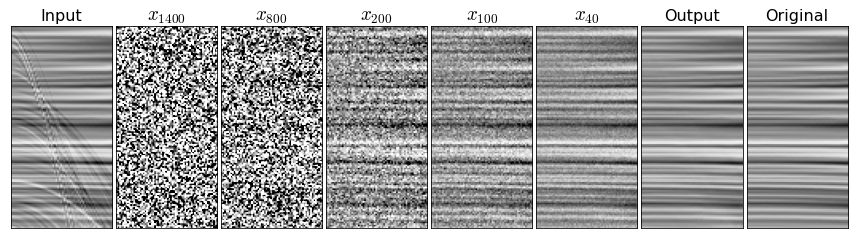}
    \end{subfigure}\\[-1ex]
    \begin{subfigure}{\textwidth}
        \centering
        \includegraphics[width=\linewidth]{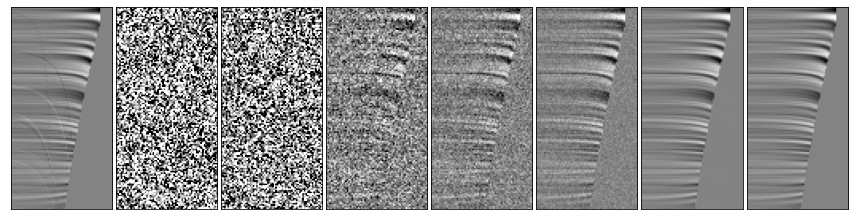}
    \end{subfigure}\\[-1ex]
    \begin{subfigure}{\textwidth}
        \centering
        \includegraphics[width=\linewidth]{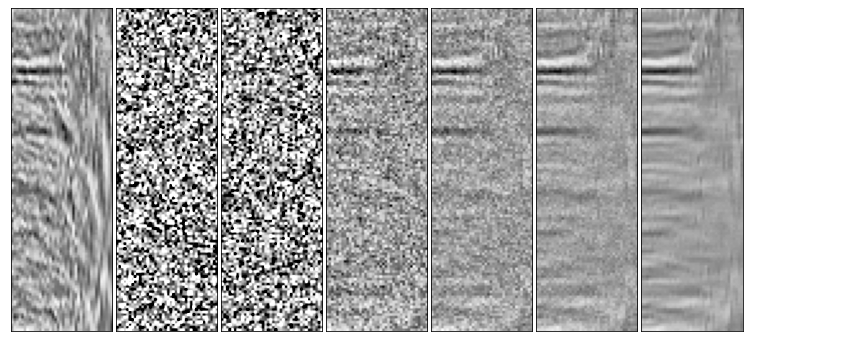}
    \end{subfigure}\\[-1ex]
    \begin{subfigure}{\textwidth}
        \centering
        \includegraphics[width=\linewidth]{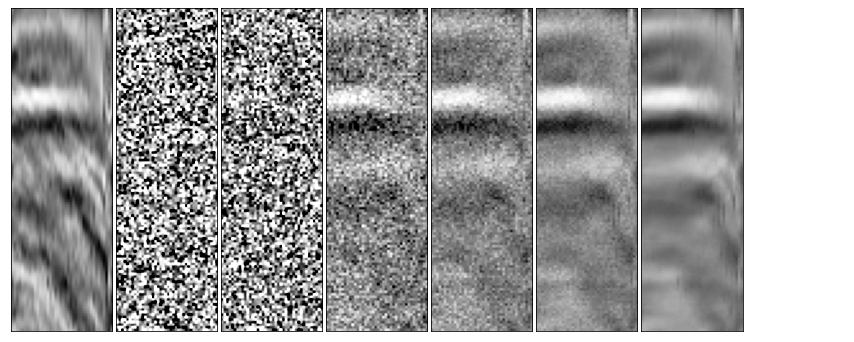}
    \end{subfigure}\\[-1ex]
    \caption{This figure displays demultiple results at different intermediate steps for the reverse process.
    Note that the first two rows show synthetic data examples, while the last two from the Volve dataset.}
    \label{fig:demo_evo}
\end{figure*}

\begin{figure*}[h]
    \begin{subfigure}{\textwidth}
        \centering
        \includegraphics[width=\linewidth]{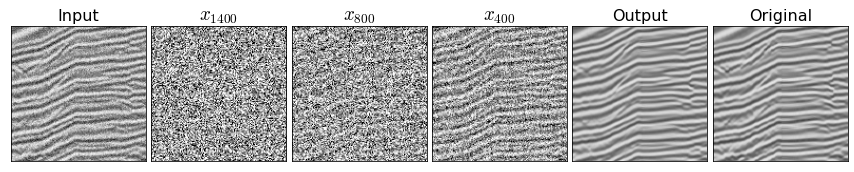}
    \end{subfigure}\\[-1ex]
    \begin{subfigure}{\textwidth}
        \centering
        \includegraphics[width=\linewidth]{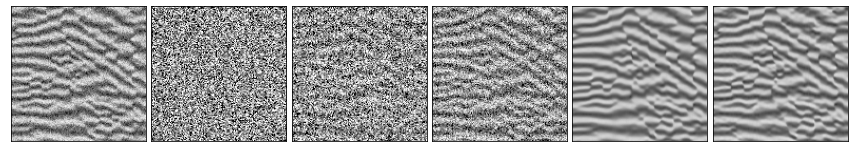}
    \end{subfigure}\\[-1ex]
    \begin{subfigure}{\textwidth}
        \centering
        \includegraphics[width=\linewidth]{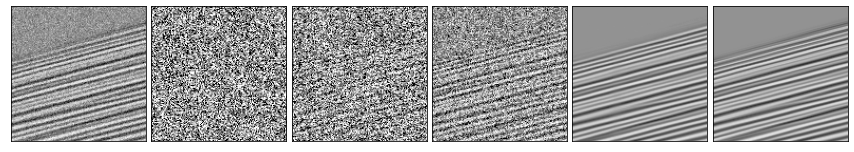}
    \end{subfigure}\\[-1ex]
    \begin{subfigure}{\textwidth}
        \centering
        \includegraphics[width=\linewidth]{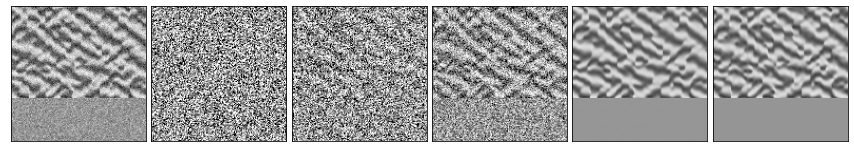}
    \end{subfigure}\\[-1ex]
    \begin{subfigure}{\textwidth}
        \centering
        \includegraphics[width=\linewidth]{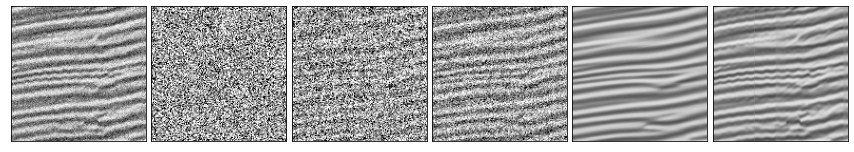}
    \end{subfigure}\\[-1ex]
    \begin{subfigure}{\textwidth}
        \centering
        \includegraphics[width=\linewidth]{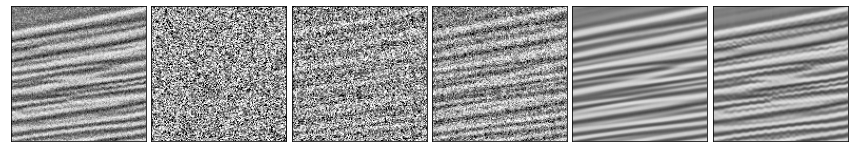}
    \end{subfigure}\\[-1ex]
    \begin{subfigure}{\textwidth}
        \centering
        \includegraphics[width=\linewidth]{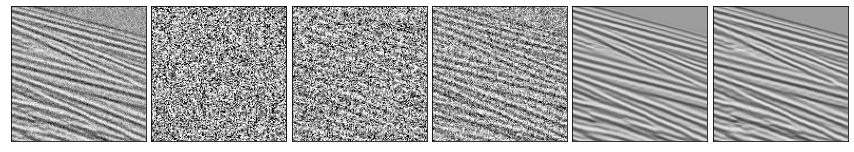}
    \end{subfigure}\\[-1ex]
    \caption{This figure displays denoising results at different intermediate steps for the reverse process.
    Note that the examples belong to the Model94 dataset.}
    \label{fig:denoi_evo}
\end{figure*}

\begin{figure*}[h]
    \begin{subfigure}{\textwidth}
        \centering
        \includegraphics[width=\linewidth]{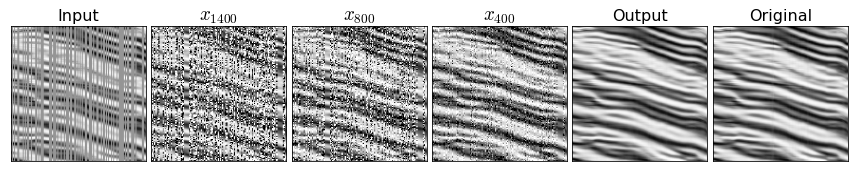}
    \end{subfigure}\\[-1ex]
    \begin{subfigure}{\textwidth}
        \centering
        \includegraphics[width=\linewidth]{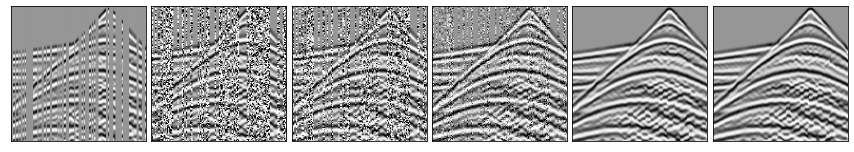}
    \end{subfigure}\\[-1ex]
    \begin{subfigure}{\textwidth}
        \centering
        \includegraphics[width=\linewidth]{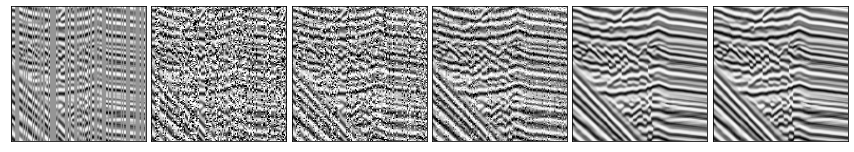}
    \end{subfigure}\\[-1ex]
    \begin{subfigure}{\textwidth}
        \centering
        \includegraphics[width=\linewidth]{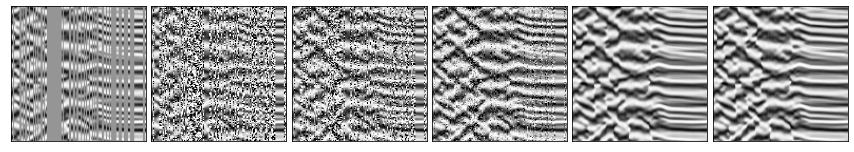}
    \end{subfigure}\\[-1ex]
    \begin{subfigure}{\textwidth}
        \centering
        \includegraphics[width=\linewidth]{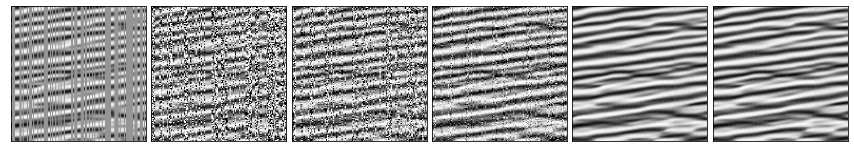}
    \end{subfigure}\\[-1ex]
    \begin{subfigure}{\textwidth}
        \centering
        \includegraphics[width=\linewidth]{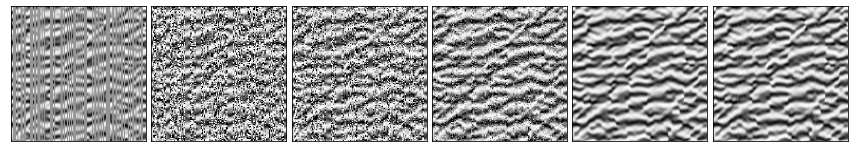}
    \end{subfigure}\\[-1ex]
    \begin{subfigure}{\textwidth}
        \centering
        \includegraphics[width=\linewidth]{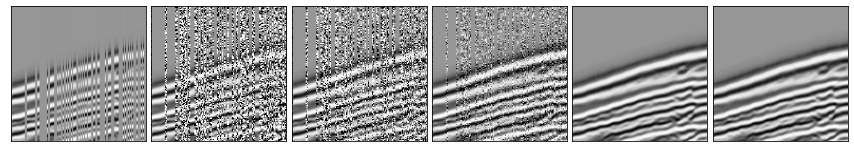}
    \end{subfigure}\\[-1ex]
    \caption{This figure displays interpolation results at different intermediate steps for the reverse process.
    Note that the examples belong to the Model94 dataset.}
    \label{fig:inter_evo}
\end{figure*}

We provide additional results, where we can visualize the evolution of the reverse process for all the aforementioned case studies (see Figure \ref{fig:demo_evo}, Figure \ref{fig:denoi_evo} and Figure \ref{fig:inter_evo}).
Note that the subindexes of the $x$ indicate the output of an intermediate step during the inference process, being $x_{1999}$ random noise and $x_{0}$ the final output of the model.

%https://www.assemblyai.com/blog/diffusion-models-for-machine-learning-introduction/
%https://developer.nvidia.com/blog/improving-diffusion-models-as-an-alternative-to-gans-part-1/
%https://angusturner.github.io/generative_models/2021/06/29/diffusion-probabilistic-models-I.html
%https://maciejdomagala.github.io/generative_models/2022/06/06/recent-rise-of-diffusion-based-models.html
%https://lilianweng.github.io/posts/2021-07-11-diffusion-models/
%https://www.youtube.com/watch?v=HoKDTa5jHvg
%https://yang-song.github.io/blog/2021/score/

\end{document}